\documentclass[a4paper]{article}

\usepackage{multicol}
\def\1{\c{c}}
\def\2{\c{C}}
\def\3{\.{I}}
\def\4{\"{a}}
\def\5{{\i}}
\def\6{$\beta$}
\def\7{\"{o}}
\def\8{\"{O}}
\def\9{\c{s}}
\def\0{\c{S}}
\def\*{\"{u}}
\def\;{\u{g}}
\def\:{\u{G}}

\usepackage{icrc2013}

\title{Modeling the Shell of Cassiopeia A to find the TeV Gamma-ray Emission Region}

\shorttitle{Modeling the Shell of Cassiopeia A to find the TeV Gamma-ray Emission Region}

\authors{
T\*l\*n Ergin$^{1}$,
Lab Saha$^{2}$,
Pratik Majumdar$^{2}$,
Mustafa Bozkurt$^{3,*}$,
E. Nihal Ercan$^{3}$.
}

\afiliations{
$^1$ TUBITAK SpaceTechnologies Research Institute, Ankara, Turkey.\\
$^2$ Saha Institute of Nuclear Physics, Kolkata, India.\\
$^3$ Bogazici University, Physics Department, Istanbul, Turkey.\\
\scriptsize{
$^{*}$ now at: Tufts University, Physics and Astronomy Department.
}
}

\email{tulun.ergin@tubitak.gov.tr}

\abstract{We will present the multi-wavelength modeling of the supernova remnant Cassiopeia A's shell based on radio, X-rays, and GeV-TeV gamma rays. Our aim is to estimate the location of TeV gamma rays with the help of spectral analysis of X-rays from different regions of the shell, because Chandra X-ray observations have a far better angular resolution than the gamma-ray measurements. Our analysis shows X-ray flux levels from various regions of the remnant to be different. We find that leptonic model is unable to explain the GeV and TeV data, simultaneously. So, we invoke a hadronic model as an additional component to explain the GeV and TeV data.}

\keywords{Supernova Remnants, Gamma Rays, X-Rays, Fermi-LAT, Chandra.}

\begin{document}
\maketitle
\begin{figure}[t]
\centering
\includegraphics[width=0.5\textwidth]{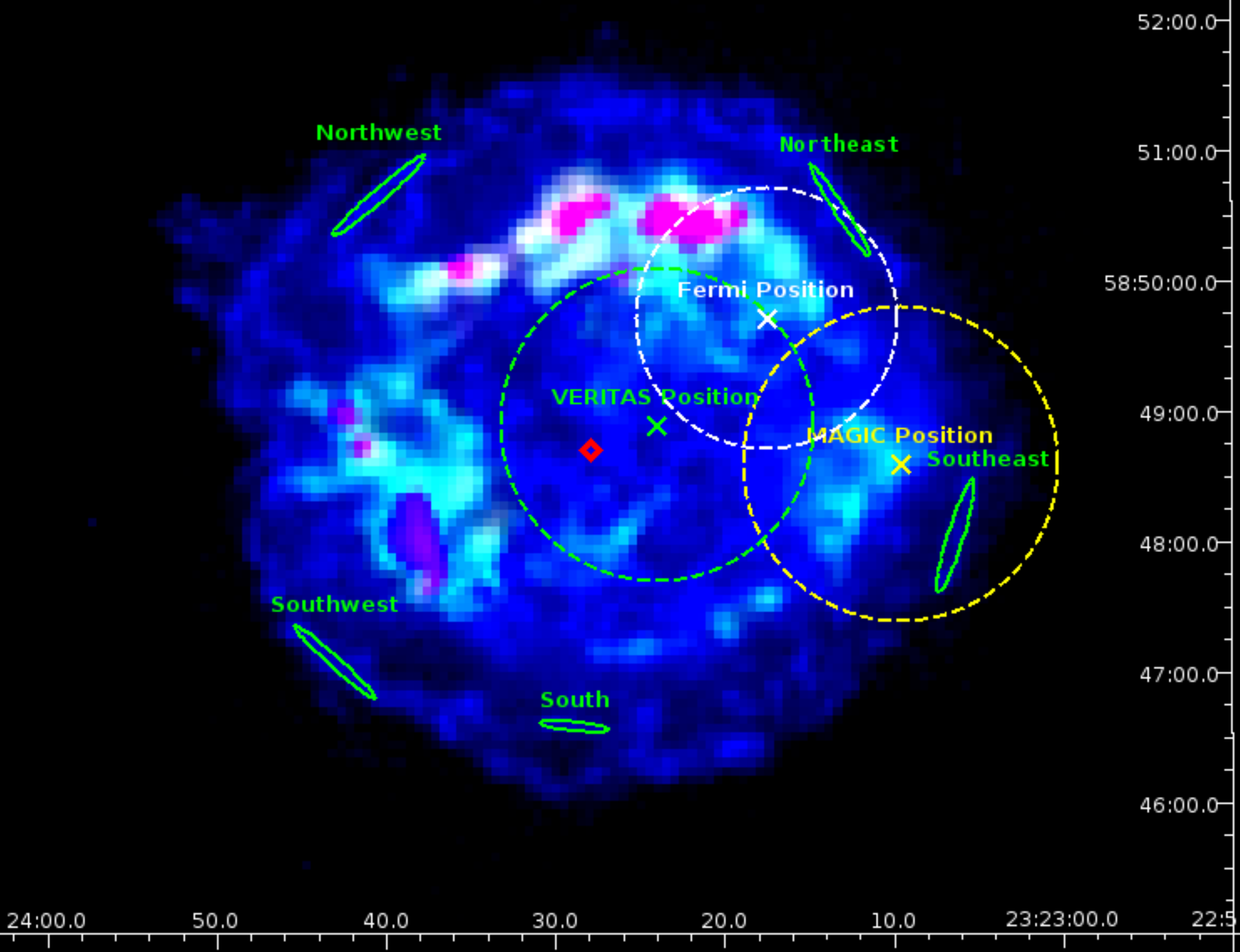}
\caption{\small{Multi-color (declination vs. right ascension in J2000) image of Cas A produced using Chandra X-ray data. The red, green, and blue color hues represent the energy ranges of [0.7,1.0], [1.0,3.5], and [3.5,8.0] keV, respectively. The red and green hues are smoothed in linear color scale, while the blue hues are shown in logarithmic scale to enhance the view of the smallest number of X-ray counts existing in the outer shell. The green ellipses represent the S, SW, SE, NW, and NE of the shell. The green, yellow, and white crosses and dashed lines correspond to the VERITAS, MAGIC, and Fermi-LAT detected gamma-ray emission locations and location error circles. The CCO location is shown with a red open diamond.}}
\label{figure_1}
\end{figure}

\section{Introduction}

Cassiopeia A (Cas A) is a historically well-known remnant observed in almost all wavebands (Radio: \cite{baars1977}, \cite{anderson1995}, \cite{vinyaikin2007}, \cite{helmboldt2009}; Optical: \cite{reed1995}; X-rays: \cite{allen1997}, \cite{maeda2009}). Cas A has been observed in gamma rays by HEGRA (\cite{aharonian2001}), MAGIC (\cite{albert2007}), and VERITAS (\cite{acciari2010}) telescopes, as well as by the Large Area Telescope (LAT) on the Fermi satellite (\cite{abdo2010}). 

The symmetric and unbroken shell structure of Cas A shows that the SNR isn't interacting with dense molecular clouds, but the short and clumpy structure of filaments on the outer shell of Cas A (as seen at the shell of SNR RXJ 1713) indicates that these filaments might be formed in the turbulent medium of interaction between SNR's shell and dense clouds, \cite{bamba2005}. Infrared observations (\cite{rho2012}) revealed CO data distributed over different parts of the shell of Cas A in varying density. Although most of the CO emission was detected from the south region of the shell, there was no maser emission detected. In X-rays, Chandra observed the shell of Cas A with high angular accuracy ($\sim$0.5$''$). Since the angular resolution is worse for gamma-ray measurements (~360$''$), Cas A was observed as a point-like source in gamma rays. There is a CCO (compact central object) located very close to the center of Cas A (\cite{pavlov2000}), which could be related to the TeV gamma-ray emission, because it position is within the error circle of the VERITAS gamma-ray source location. 

Figure \ref{figure_1} shows the Chandra X-ray image of Cas A between energies 0.7 and 8 keV, where the blue tones are the highest energy counts (3.5$-$8 keV), while red and green tones are the lower energy ranges of 0.7$-$1.0 and 1.0$-$3.5 keV, respectively. The selected regions contain filaments dominated by non-thermal emission (\cite{yamazaki2003}, \cite{bamba2005}, \cite{araya2010}, \cite{araya2012}), which mostly shine in X-ray energies between 3.5 and 8 keV. 
\section{Data Analysis and Results}
We have analyzed the Chandra X-ray data of Cas A by focusing the analysis on the selected filaments from the northwest (NW), northeast (NE), south (S), southwest (SW), and southeast (SE) of the shell (\cite{bozkurt2013}). These regions are represented by green ellipses on the color-coded Figure \ref{figure_1}.
The locations (with green, yellow, and white crosses) and location errors (with green, yellow, and white dashed circles) of the TeV and GeV gamma-ray emissions as measured by VERITAS, MAGIC, and Fermi-LAT, respectively, are also shown on Figure \ref{figure_1}. TeV locations found by VERITAS and MAGIC are more toward the east and southeast of the shell, while the GeV location of Fermi-LAT is toward the northeast of the remnant. But we note here that the point-spread function of a point-like source of all three detectors is bigger in comparison to the radio size of the shell (5$'$). Therefore, it is not really clear at what part of the shell the GeV and TeV gamma-ray emission dominates.  

The power-law model was fit to X-ray spectrum for each region and the emission lines of iron (Fe), silicon (Si), sulphur (S) and other elements were fit with Gaussian functions. For each region, we obtained the following fit parameters: Spectral index and spectral normalization. These two parameters are used for calculating each region's flux. 
So, the X-ray fluxes of the S, SW, SE, NW, and NE regions are shown on Figure \ref{figure_1} with red, green, blue, brown, and black thick stripes, respectively. We have also analyzed the GeV gamma-ray data of Fermi-LAT, where the details of this analysis are explained in \cite{bozkurt2013}. The spectral data points of the GeV emission obtained from this analysis are shown as green boxes in Figures \ref{figure_2} and \ref{figure_3}. To model the multi-waveband spectrum, we also included the MAGIC spectral data points (\cite{albert2007}) shown with black filled circles and error bars in Figures \ref{figure_2} and \ref{figure_3}.
\section{Modeling the Spectrum}
\subsection{Leptonic Model}
The non-thermal X-ray emission in the selected regions can be explained by the synchrotron emission. The flux of the synchrotron emission (F$_{synch}$) depends on the magnetic field strength (B).  $$\mbox{F}_{synch} \propto \mbox{B}^{p+1}~~~,$$ where p is the spectral index of the power-law type spectrum of the high energy electrons distributed over the whole remnant. The open form of the F$_{synch}$ function is given in Appendix B of \cite{zirakashvili2010} or in \cite{kelner2006}. From the observed radio spectrum, $S_{\nu}  \propto \nu^{-0.77}$ (\cite{baars1977}), we calculated the power-law spectral index, p, for electron distribution to be 2.54. 

\begin{figure}[t]
\centering
\includegraphics[width=0.5\textwidth]{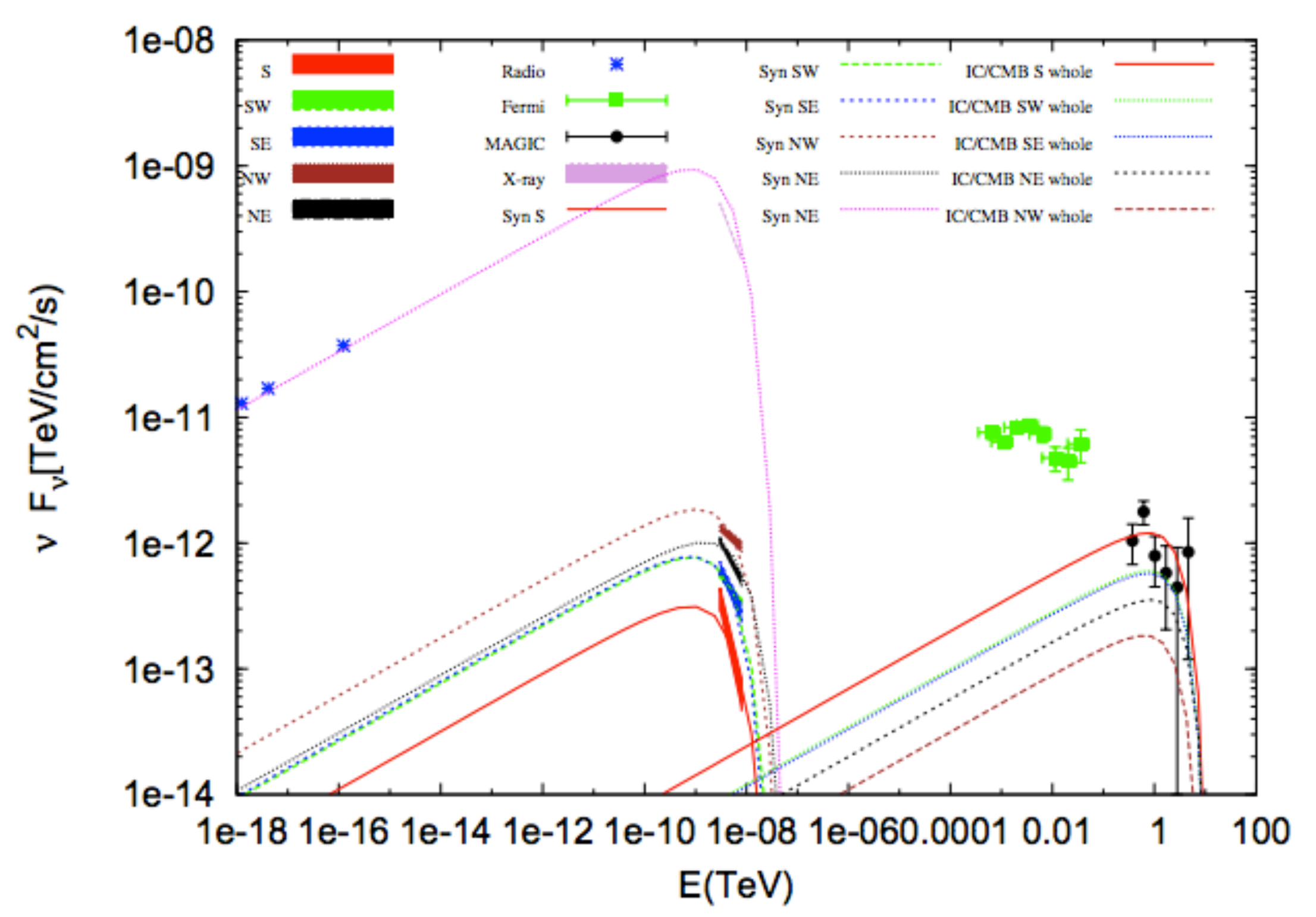}
\caption{Synchrotron and IC emission spectra along with the data for different regions of the shell.}
\label{figure_2}
\end{figure}

The value of $\mbox{F}_{synch}$ changes with the electron density ($\rho_e$) in the environment. In the two-zone scenario of Cas A, the magnetic field energy densities are different for two different zones, but the density of relativistic electrons in these zones are comparable to each other, \cite{atoyan2000}. Hence, we use the uniform electron density for all the shell regions. If we express the X-ray flux with the following relation $$\mbox{F}_{synch} = \mbox{A E}^{-p}~~~,$$ where p is the spectral index and A is the normalization parameter, we can explain the differences in the X-ray fluxes for different regions by the differences in the normalization and spectral index parameters. Therefore, the contribution to the X-rays from these regions is not the same and this leads to different levels of gamma rays from different regions. Also, taking the density of the electrons as a constant, we can estimate the magnetic field strength, B. Since F$_{synch}$ was found to be different for every region, we ended up with different B values for each region. 
\begin{table}[h]
\begin{center}
\begin{tabular}{|l|c|}
\hline Region              & Magnetic Field (B) [$\mu$G]   \\ \hline
           South                 & 90                                                \\ \hline
           Southwest        & 120                                              \\ \hline
           Southeast         & 120                                              \\ \hline
           Northwest         & 170                                              \\ \hline
           Northeast         & 150                                               \\ \hline
\end{tabular}
\caption{The magnetic field parameters for the synchrotron spectra for all selected regions. The rest of the parameters are fixed for all regions, which are the following: spectral index = 2.54, $\gamma_{min}$ = 1.0, $\gamma_{max}$ = 5.5$\times$10$^7$, distance = 3.4 kpc, normalization constant = 2.0$\times$10$^{53}$ TeV$^{-1}$. }
\label{table_1}
\end{center}
\end{table}
Here, we wanted to find the region, which contributes the most to the gamma-ray flux induced by leptonic processes. To do so, we first considered that the whole remnant is uniform in X-ray flux and the southern part of the remnant contains only a fraction of the flux from the whole remnant. Considering magnetic field to be 90 $\mu$G for the S region, we fitted the corresponding observed X-ray data and estimated the corresponding model parameters for the synchrotron emission process. Afterwards, using the same electron density and same power-law spectral index we estimated the magnetic fields for all the regions shown in Table \ref{table_1}. The fitted X-ray spectra for different regions are shown in Figure \ref{figure_2} as a red solid line for the S region, and green, blue, brown, and black dashed lines for the SW, SE, NW, and NE regions, respectively. 

We then scale those spectra to the whole remnant flux and obtain the corresponding scaling factors for those regions. So, we obtain the electron distribution parameters for every selected region using the corresponding X-ray flux after scaling it to the whole remnant's flux. The parameters are shown in Table \ref{table_2}. From the leptonic model the total energy of leptons was estimated to be E$_e$ = 3.8$\times$10$^{50}$ ergs. Using those parameters we obtain the inverse Compton (IC) spectra considering cosmic microwave background photons as target photons for this emission process. 
Based on the flux upper limit given by SAS-2 ad COS B detectors, a lower limit on the magnetic field in the shell of Cas A was estimated to be 8$\times$10$^{-5}$G and also to be consistent with the magnetic field B = 80 $-$160 $\mu$G estimated by \cite{vink2003} we considered magnetic field 90 $\mu$G for the S region. The magnetic field for other regions is estimated with respect to the magnetic field considered in the S region. If the shell region is dominated by strong magnetic field (e.g. 100 $\mu$G), then the IC component of radiation is reduced. According to our analysis, we found that southern part of the shell is dominant in producing IC among all the regions. 
\begin{table}[h]
\begin{center}
\begin{tabular}{|l|c|}
\hline Parameters                           & Values                                                   \\ \hline
           $\gamma_{min}$                 & 1                                                             \\ \hline
           $\gamma_{max}$                & 5$\times$10$^7$                                \\ \hline
           n$_H$                                    & 10 cm$^{-3}$                                       \\ \hline
           $\alpha$                                & 2.54                                                        \\ \hline
           N                                             & 2.6$\times$10$^{56}$   TeV$^{-1}$ \\ \hline\hline\
           Total                                       & 3.8$\times$10$^{50}$ ergs               \\ \hline
\end{tabular}
\caption{The Bremsstrahlung process parameters for the S region of the SNR.}
\label{table_2}
\end{center}
\end{table}
The lines going through the MAGIC and Fermi-LAT data points on Figure \ref{figure_2} are the estimated fluxes from the IC scattering calculated for the S (red), SW (green), SE (blue), NW (brown), and NE (black) regions. It seems as if the TeV data overlaps better with the IC emission prediction for the south of the remnant. But TeV spectral data points at the highest energy bins also fits with the IC emission function estimated for the SW or SE regions. Additionally, the Fermi-LAT spectral data points at GeV energies can't be explained by the IC mechanism alone and it has to be modeled by either the Bremsstrahlung process or with neutral pion decay. 

\begin{figure}[t]
\centering
\includegraphics[width=0.5\textwidth]{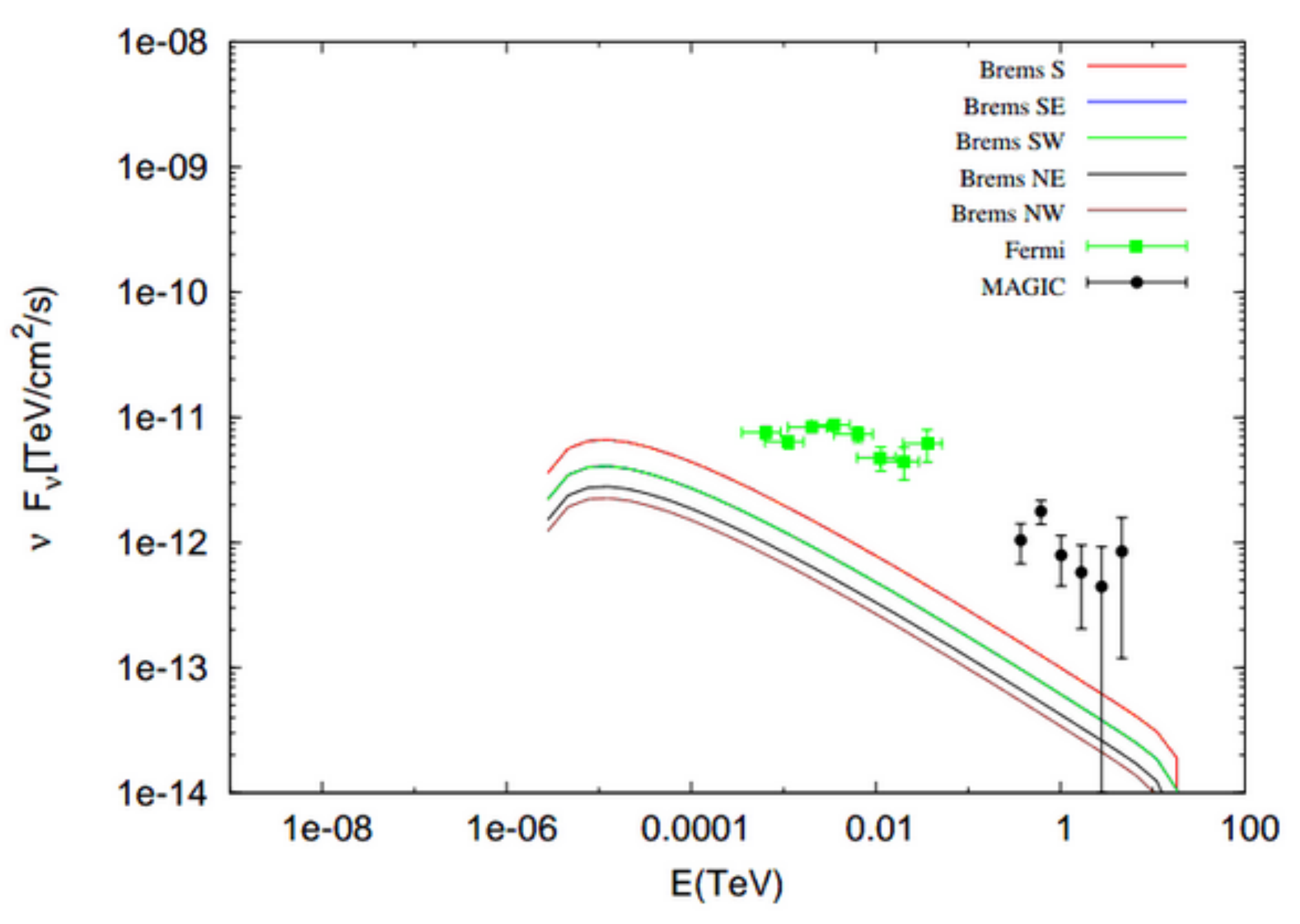}
\caption{Bremsstrahlung spectra for n$_{\mbox{\footnotesize{H}}}$ = 10 cm$^{-3}$ and for the S, SE, SW, NE, and NW of the shell (red, blue, green, and black solid lines, respectively).}
\label{figure_3}
\end{figure}

We also estimated the contribution of Bremsstrahlung process to the TeV energies. We used the parameters for the S region as given in Table \ref{table_2} to calculate Bremsstrahlung spectrum. From the red, blue, green, and black solid lines on Figure \ref{figure_3} representing the S, SE, SW, NE, and NW regions of the SNR, it is obvious that Bremsstrahlung process cannot explain the TeV data.  The Bremsstrahlung flux linearly depends on the ambient proton density. For our estimations we considered the ambient proton density to be n$_H$=10 cm$^{-3}$ (\cite{laming2003}). The Bremsstrahlung process can explain TeV data only for higher values of ambient proton density, which is about 70 cm$^{-3}$. This high density of ambient protons is unusual for a remnant unless there are molecular clouds in this region. 
As we know, about 10\% of the explosion energy of SNR is converted to the energy of relativistic particles and the total explosion energy of supernovae is about 10$^{51}$ ergs. Using M$_{ejecta}$ = 2M$_{\odot}$ (\cite{willingale2003}, \cite{laming2003}), Fermi-LAT collaboration calculated the effective gas density as n$_{eff}$ $\simeq$ 32 cm$^{-3}$, \cite{abdo2010}. Also, recently \cite{rho2012} reported that southern part of the remnant shows presence of high-density CO molecules. 

\subsection{Hadronic Model}
We also calculated the gamma-ray flux resulting from the pion ($\pi^{0}$) decay process. For this purpose, we first selected the S region, because as seen on the right side of Figure \ref{figure_2} the largest amount of gamma-ray flux seems to be coming from this region. 

\begin{figure}[t]
\centering
\includegraphics[width=0.5\textwidth]{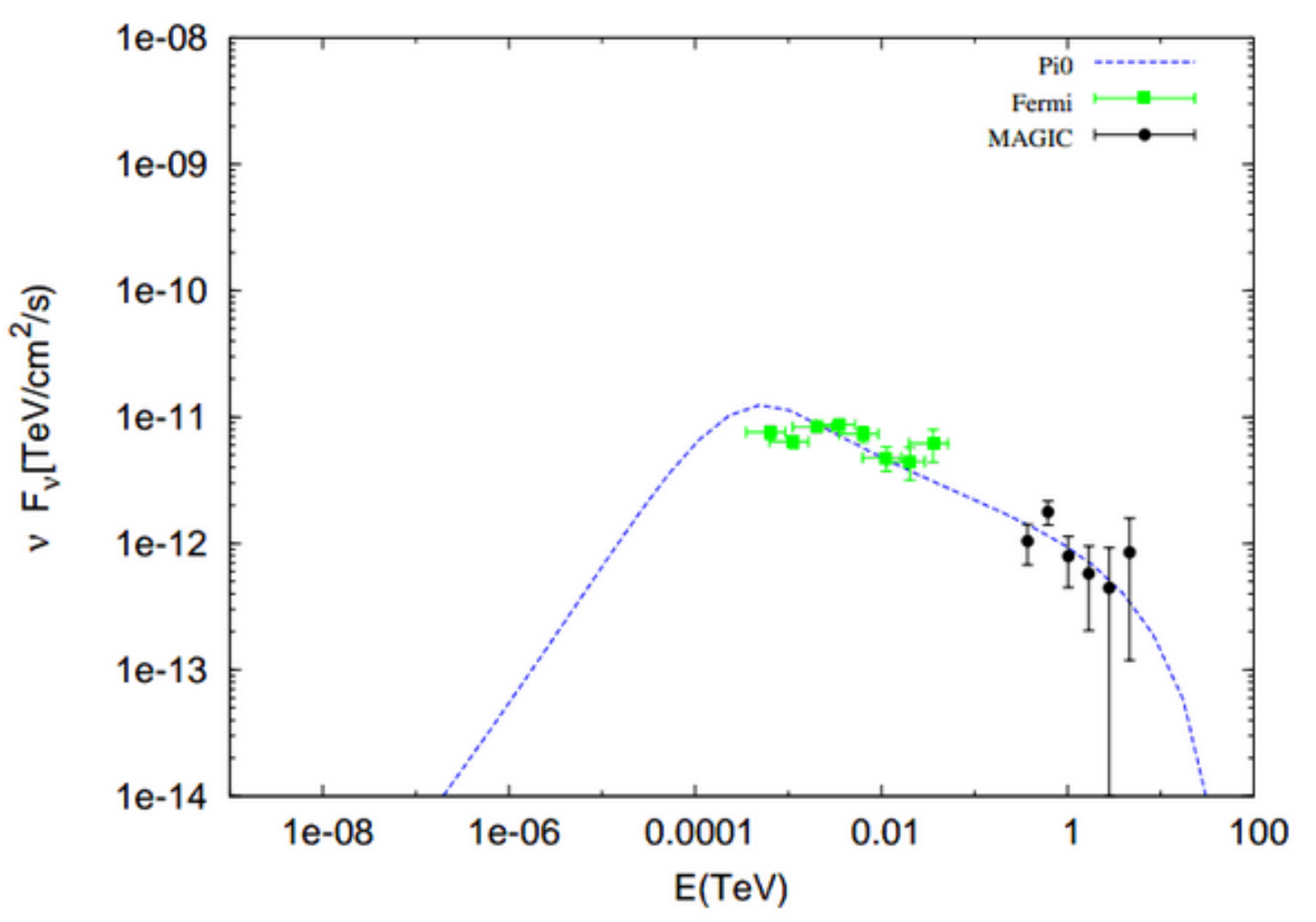}
\caption{$\pi^{0}$ decay spectra for power-law proton spectra for n$_{\mbox{\footnotesize{H}}}$ = 10 cm$^{-3}$ (blue dashed line).}
\label{figure_4}
\end{figure}

For hadronic contribution to the gamma-ray flux through the decay of $\pi^{0}$, we considered ambient proton density to be 10 cm$^{-3}$. Accelerated proton spectrum was taken as dN/dE$_p$ $\propto$ E$^{-2.35}$ with an exponential cutoff at 80 TeV. Figure \ref{figure_4} shows the contribution of gamma-ray flux from the $\pi^{0}$ decay calculated by following \cite{kelner2006}. The gamma-ray spectrum was fitted within the observed GeV$-$TeV energy range and the corresponding parameters are shown in Table \ref{table_3}. Taking n$_H$ = 10 cm$^{-3}$, the total energy of the protons in hadronic model is E$_p$ = 3.9$\times$10$^{48}$ ergs. 
\begin{table}[h]
\begin{center}
\begin{tabular}{|l|c|}
\hline Parameters             & Values					        \\ \hline
           $E_{cut}$                & 80 TeV                                                 \\ \hline
           n$_H$                      & 10 cm$^{-3}$                                     \\ \hline
           $\alpha$                  & 2.35                                                      \\ \hline
           N                               & 2.0$\times$10$^{48}$ TeV$^{-1}$ \\ \hline\hline\
           Total                         & 3.9$\times$10$^{48}$ ergs             \\ \hline
\end{tabular}
\caption{The neutral pion decay process parameters for the S region of the SNR.}
\label{table_3}
\end{center}
\end{table}

Although increasing the effective density of the ambient gas to higher values (than the estimated average density) may help the Bremsstrahlung model to better fit to the GeV$-$TeV data, the gamma-ray flux due to the $\pi^{0}$ decay of accelerated hadrons also increases. So, the $\pi^{0}$ decay process can't explain the GeV$-$TeV data any more, unless the total energy budget of the protons is reduced. That in turn indicates a lower conversion efficiency of the explosion energy of Cas A into accelerating protons. 

\section{Conclusion}
As a result, we found that the predicted gamma-ray emission from the IC process in the S region of the shell has the highest flux value and this predicted flux matches better to the TeV data in comparison to flux predictions from other regions of the shell. Also, the GeV and TeV gamma-ray data fits reasonably well to the hadronic model, which is independent of the selected region on the shell. The second best fitting IC prediction with the TeV data is from the SW and SE, and then NE region. If the SNR would be perfectly symmetric in shape, we would expect that the fluxes of each region of the shell should be equal. But apparently, the shell's emission is not homogeneously distributed. The reason for the variations in X-ray and gamma-ray fluxes can be due to different amounts of particles or variations in the magnetic field at different regions of the SNR's shell. Also the molecular environment might be different at different sides of the shell.  So, we need to work more on the model to minimize the discrepancies between the data and the model.

\end{document}